\documentclass[prl,superscriptaddress,twocolumn,showpacs]{revtex4}
\usepackage{bm}
\usepackage{graphicx}
\begin{document}

\newcommand{\mwimp}{$m_\chi$}
\newcommand{\sigmapsi}{$\sigma_p^{SI}$}
\newcommand{\sigmapsd}{$\sigma_p^{SD}$}
\newcommand{\sigmansd}{$\sigma_n^{SD}$}
\newcommand{\kms}{km~s$^{-1}$}
\newcommand{\vsun}{$v_\odot$}
\newcommand{\ve}{$v_\oplus$}
\newcommand{\vescsun}{$v^{\odot}_{esc}$}
\newcommand{\vesce}{$v^{\oplus}_{esc}$}

\title{Getting the astrophysics and particle physics of dark matter out of next-generation direct detection experiments}
\author{Annika H.\ G.\ Peter}
\email{apeter@astro.caltech.edu}
\affiliation{California Institute of Technology, Mail Code 249-17, 
  Pasadena, CA 91125, USA}

\begin{abstract}
The next decade will bring massive new data sets from experiments of the direct detection of weakly interacting massive particle (WIMP) dark matter.  Mapping the data sets to the particle-physics properties of dark matter is complicated not only by the considerable uncertainties in the dark-matter model, but by its poorly constrained local distribution function (the ``astrophysics'' of dark matter).  I propose a shift in how to think about direct-detection data analysis.  I show that by treating the astrophysical and particle-physics uncertainties of dark matter on equal footing, and by incorporating a combination of data sets into the analysis, one may recover both the particle physics and astrophysics of dark matter.  Not only does such an approach yield more accurate estimates of dark-matter properties, but it may illuminate how dark matter coevolves with galaxies.
\end{abstract}


\maketitle

Dark matter makes up $\sim 23\%$ of the energy density of the observable Universe, yet its identity is unknown (e.g., \cite{komatsu2010}).  While microlensing constraints limit the fraction of dark matter that consists of compact objects such as primordial black holes \cite{alcock1998}, there is a plethora of viable dark-matter particle candidates, most arising naturally in extensions to the standard model (SM) of particle physics \cite{dodelson1994}.

If the dark-matter particle is massive ($m_\chi \gtrsim 1$ GeV) and interacts with baryons, it may be observed in direct-detection experiments \cite{goodman1985}.  This class of particle is generically called a weakly-interacting massive particle (WIMP), and the experiments look for the small ($\sim 10-100$ keV) kinetic-energy transfer from a WIMP to a nucleus.  Current limits on the WIMP-nucleon spin-independent and WIMP-neutron spin-dependent cross sections come from $\sim 10$ kg-sized experiments \cite{angle2008,cdms2009}.  Ton-scale experiments are expected by $\sim 2015$, and there are plans for $\sim 10$-ton experiments by 2020 \cite{gaitskell2008}.  These experiments are touted as being able to probe far into supersymmetric and Kaluza-Klein parameter space.

The event rate in these experiments not only depends on the particle-physics properties ($m_\chi$, cross sections) of the dark matter, but on the local dark-matter distribution function (DF).  The event rate (per kg of detector mass in which the target nucleus has atomic number $N$) is
\begin{eqnarray}
	\frac{dR}{dQ} = \left( \frac{m_N}{\mathrm{kg}} \right)^{-1}  \int_{v_{\mathrm{min}}} d^3 \mathbf{v} \frac{d \sigma_N}{d Q} v f(\mathbf{x},\mathbf{v}), \label{eq:drdq}
\end{eqnarray}
where $d\sigma_N/dQ$ is the differential scattering cross section, $f(\mathbf{x},\mathbf{v})$ is the local WIMP DF,
\begin{eqnarray}
v_{\mathrm{min}} = ( m_N Q / 2 \mu^2_N)^{1/2} \label{eq:vmin}	
\end{eqnarray}
is the minimum speed required for a WIMP of mass $m_\chi$ to deposit energy $Q$ to a nucleus of mass $m_N$, and $\mu_N$ is the particle-nucleus reduced mass.  Thus, any limits or constraints on the WIMP particle-physics properties derived from the data depend on the WIMP DF.

However, the local WIMP DF is poorly understood.  Even if the local WIMP DF is dominated by a smooth, equilibrium dark-matter halo, and the density profile could be determined (e.g., Ref. \cite{strigari2009}), the velocity distribution cannot be uniquely determined \cite{binney2008}.  Dark-matter experiments are the \emph{only} probe of the local WIMP velocity distribution.  Moreover, the smooth halo may not dominate the direct-detection event rates.  Recent work has suggested that the Milky Way may also have a disk of dark matter, and it may dominate the direct-detection signal \cite{read2009}.  However, its properties depend sensitively on the accretion history of the Milky Way, which is unknown.  In addition, the WIMP velocity distribution may be dominated by a few tidal streams.  Simulations cannot currently resolve such small-scale structures.  Without better understanding of the astrophysical properties of the local dark-matter distribution, particle-physics constraints on dark matter from direct-detection experiments are essentially meaningless.

While there has been some attention to how different DFs change exclusion curves in the case of no signal, there has been little work on how to determine WIMP particle-physics properties if at least one direct-detection experiment sees a WIMP signal \cite{green2003,strigari2009}.  The bulk of that work has the underlying assumption that the direct-detection signal is dominated by the smooth halo.

New ideas and methods are required to properly extract particle-physics implications from upcoming data sets since it is not feasible to model the astrophysical properties of WIMPs.  I propose treating the astrophysical uncertainties in the WIMP DF on par with the particle-physics uncertainties, in order to derive robust constraints on \emph{both} the WIMP DF and particle physics from the data.  However, there are degeneracies among the particle-physics and astrophysics properties.  One must analyze multiple data sets together to break the degeneracies.  This approach is inspired by cosmological data analysis (e.g., \cite{komatsu2010}).

As a test of principle of how both particle-physics and astrophysics properties may be derived by a combined analysis of multiple data sets if WIMPs are detected, I perform a simple toy study employing Fisher-matrix forecasting for toy models of upcoming direct-detection experiments.  Future work will use Monte Carlo simulations for forecasting instead of Fisher matrices, but Fisher matrices suffice for this first look into how well both the particle-physics and astrophysics properties of WIMPs can be inferred from future data sets.  For this study, the local WIMP DF is treated as a Maxwellian with a one-dimensional velocity dispersion $v_{\mathrm{rms}}$, with the Solar System lagging the WIMP distribution by a velocity $v_{\mathrm{lag}}$.  I place no restrictions on $v_{\mathrm{rms}}$ or $v_{\mathrm{lag}}$, in other words, there are flat priors on the velocity parameters.  Unlike Ref. \cite{strigari2009}, there is no prior that the direct-detection event rate must be dominated by smooth halo WIMPs.  I only assume that for whatever WIMP structure dominates the event rates, the velocity distribution is described by a single Maxwellian distribution.  I use this simple form of the velocity distribution without priors on the parameters because this work is a test of principle, to show the power of a joint analysis without the theoretical prior on which WIMP population dominates the event rates.  In later work I will input more complex velocity distributions and extract the velocity distribution in a nonparametric way, perhaps akin to Ref. \cite{drees2007}.   

The free parameters are $m_\chi$; $D = \rho_\chi \sigma_p^{SI} / (2 m_p^2)$, the normalization of the differential event rate, where $\rho_\chi$ is the local dark-matter density, and $\sigma_p^{SI}$ is the spin-independent WIMP-proton cross section; $v_{\mathrm{lag}}$; and $v_{\mathrm{rms}}$.  I set $\rho_\chi = 0.3\hbox{ GeV cm}^{-3}$ for concreteness, although the exact value of the WIMP density is not relevant for this analysis, as $\rho_\chi$ is degenerate with $\sigma_p^{SI}$.  I set the spin-dependent cross sections to zero, a restriction I will loosen in future work.  I use toy models for a set of direct-detection experiments with data sets in 2015 and 2020.  These experiments can distinguish nuclear-recoil events from backgrounds on an event-by-event basis, but they cannot determine the recoil direction (although dark-matter time projection chambers can \cite{sciolla2008}).  The 2015-era set of experiments I use are as follows: (1) A xenon-based experiment inspired by XENON1T \cite{aprile2009b} with a nuclear recoil analysis window $2-27$ keV and $\sim 1/6$ efficiency of finding nuclear recoil events in the 1-ton experimental volume over the course of one year.  (2) A xenon-based experiment inspired by LUX with a xenon mass of 300 kg \cite{gaitskell2008}.  I assume a three-year exposure with an analysis window $5-27$ keV and a $1/6$ efficiency.  (3) A germanium experiment modeled on SuperCDMS Phase B.  I assume a one-year exposure, 145-kg target mass, and an efficiency and analysis window similar to that of CDMS-II ($10-100$ keV) \cite{akerib2008,cdms2009}.  (4) An argon experiment inspired by WArP, with a target mass of 140 kg, threshold of 40 keV, three-year exposure, and efficiency similar to that of the prototype.  The larger version of WArP is different in form from the prototype, so the efficiency and analysis window of the actual WArP experiment is likely to be different than what I assume here \cite{szelc2009}.  The 2020-era experiments used are as follows: (1) A 20-ton xenon experiment, whose characteristics I model like the LUX-inspired experiment.  (2) A 1-ton germanium experiment with characteristics of the 2015-era germanium experiment.  (3) A 10-ton argon experiment.  All 2020-era experiments are assumed to have one-year exposures.

I perform the analysis assuming $m_\chi = 100$ GeV, $\sigma_p^{SI} = 10^{-44} \hbox{ cm}^2$ ($D = 1.4 \times 10^{-45}\hbox{ cm}^{-1}\hbox{ GeV}^{-1}$), $v_{\mathrm{lag}} = 220 \hbox{ km s}^{-1}$, and $v_{\mathrm{rms}} = 155\hbox{ km s}^{-1}$.  Error ellipses are based on Poisson likelihoods,
\begin{eqnarray}
	\mathcal{L} = \frac{N_e^{N_o}}{N_o!}e^{-N_e} \prod_{i=1}^{N_o} \mathcal{L}_i(Q_o^i|\theta),
\end{eqnarray}
where $N_e$ is the expected number of events given the theoretical parameters $\theta$, $N_o$ is the observed number of events with observed energies ($Q_o^1,Q_o^2,...,Q_o^{N_o}$), and $\mathcal{L}_i$ is the likelihood that a recoil $i$ with observed energy $Q_o^i$ is found given the theoretical parameters.  The likelihood for each individual event is
\begin{eqnarray}
	\mathcal{L}_i = \frac{\mathcal{E}(Q_o^i) \frac{dR}{dQ}(Q_o^i)}{\int dQ \mathcal{E}(Q) \frac{dR}{dQ}(Q)},
\end{eqnarray}
where $\mathcal{E}$ is the efficiency of identifying an event with energy $Q_o^i$.  For recoils outside the analysis window, $\mathcal{E}(Q_o^i) = 0$.  For now, I neglect energy errors.  The analysis of the germanium experiments with the CDMS energy errors described in Ref. \cite{cdms2009} is virtually indistinguishable from the case of perfect energy resolution.

\begin{figure}
	\begin{center}
	\includegraphics[width=0.45\textwidth]{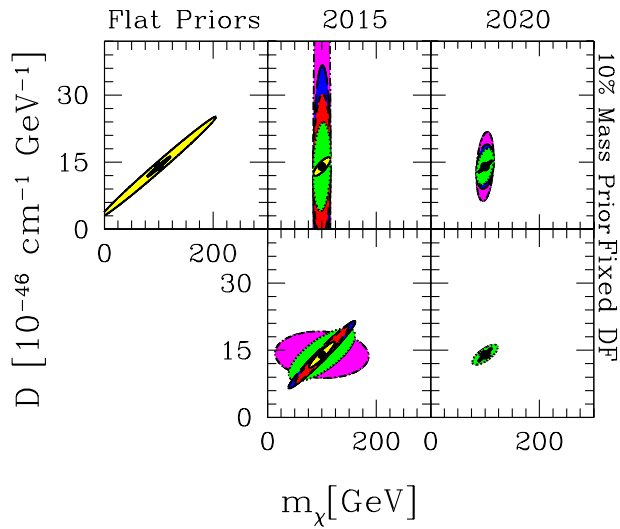}
	\end{center}
	\caption{\label{fig:mA}$1\sigma$ error ellipses for $m_\chi$ and $D$ if $m_\chi = 100$ GeV, \sigmapsi$=10^{-44}\hbox{ cm}^2$, and $\rho_\chi = 0.3\hbox{ GeV cm}^{-3}$.  \emph{Upper left panel:} Forecasted errors for 2015-era (\emph{yellow or lightest grey}) and 2020-era (\emph{black}) experiments with flat priors on all parameters.  \emph{Upper center panel:} Error ellipses assuming the LHC has measured the WIMP mass to 10\%, for 2015-era experiments.  The ellipses represents errors obtained with the WArP-inspired (magenta; background medium grey), LUX-inspired (blue; dark grey), XENON1T-inspired (red; foreground medium grey), and SuperCDMS-inspired (green; light grey) experiments.  The yellow (lightest grey) ellipse shows errors for the joint analysis of all experiments with the LHC mass prior. \emph{Upper right panel:} Error ellipses with the LHC mass prior for 2020-era experiments (blue or dark grey: xenon, green or light grey: germanium experiment, and magenta or medium grey: argon).  The diagonal line shows the errors obtained in a joint analysis of all experiments with the LHC mass prior. \emph{Lower panels:} Error ellipses if the WIMP DF is fixed.}
\end{figure}

\begin{figure}
	\begin{center}
	\includegraphics[width=0.45\textwidth]{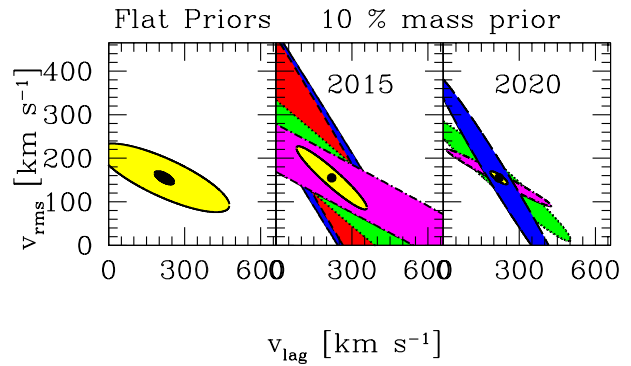}
	\end{center}
	\caption{\label{fig:vel}$1\sigma$ errors in the velocity parameters.  \emph{Left panel:} Errors obtained in a joint analysis of all 2015-era (yellow or lightest grey) and 2020-era (black) experiments with flat priors on all parameters.  \emph{Center panel:} Error ellipses for each 2015-era experiment analyzed with an LHC mass prior.  The yellow (lightest grey) ellipse shows the errors for the joint analysis.  \emph{Right panel:} Errors for the 2020-era experiments and an LHC mass prior.  Experiments the same as for Fig. \ref{fig:mA}. }
\end{figure}

In Fig.~\ref{fig:mA}, I show $1\sigma$ errors in the $m_\chi$-$D$ plane.  In the left-hand plot, I show errors assuming flat priors in all parameters, meaning that there is no external information on parameters from other experiments or observations.  The lightly filled ellipse corresponds to the 2015-era experiments, and the dark ellipse corresponds to the 2020-era experiments.  In Fig.~\ref{fig:vel}, the left-hand plot shows errors in the $v_{\mathrm{lag}}$-$v_{\mathrm{rms}}$ plane with flat priors.  While the 2015-era experiments do not constrain \mwimp, $D$, or $v_{\mathrm{lag}}$, $v_{\mathrm{rms}}$ can be determined to $\sim$50\%.  The 2020-era experiments can constrain all four parameters to of order 20$\%$.   

I compare these errors against those found either with a mass prior or assuming that the local WIMP DF is perfectly known.  In the upper central and right-hand panels of Fig. 1, I show error ellipses assuming that the LHC has measured \mwimp~to 10\%, which is the expected uncertainty in \mwimp~if the LHC identifies a supersymmetric WIMP \cite{baltz2006}.  All parameters but $m_\chi$~have flat priors.  While there are degeneracies among parameters in each individual experiment, the experiments complement each other to drive down the uncertainties, as does the mass prior.  In the lower central and right-hand panels in Fig. 1, I show the errors assuming that the local WIMP DF is known exactly.  The errors are deceptively small in the 2015 forecast, but the errors are not much smaller in the 2020-era experiments.  However, in both cases, a fixed local WIMP DF will yield significant biases in $m_\chi$ and $D$ as well as too-small error bars \cite{strigari2009}.

In Fig.~\ref{fig:vel}, I show $1\sigma$ uncertainties in the velocity parameters.  As in Fig.~\ref{fig:mA}, I show errors for each experiment analyzed with the LHC mass prior for the 2015- and 2020-era experiments, as well as the errors if all experiments are analyzed together with the mass prior.  The broad degeneracy between $v_{\mathrm{lag}}$ and $v_{\mathrm{rms}}$ in each experiment is broken since the direction of the degeneracy is different in each experiment.  In the joint analysis, the uncertainties in the velocity parameters are only slightly reduced with the LHC mass prior than if \mwimp~were a free parameter.

\begin{figure}[t]
	\begin{center}
	\includegraphics[width=0.45\textwidth]{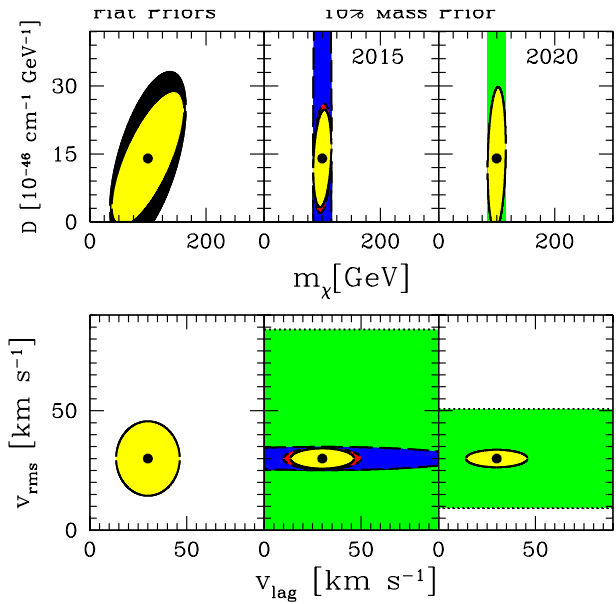}
	\end{center}
	\caption{\label{fig:low}Same particle-physics parameters as Figs. \ref{fig:mA} and \ref{fig:vel} but with $v_{\mathrm{lag}} = v_{\mathrm{rms}} = 30\hbox{ km s}^{-1}$.  Ellipse colors have the same meanings as before.}
\end{figure}

\begin{figure}[t]
	\begin{center}
	\includegraphics[width=0.45\textwidth]{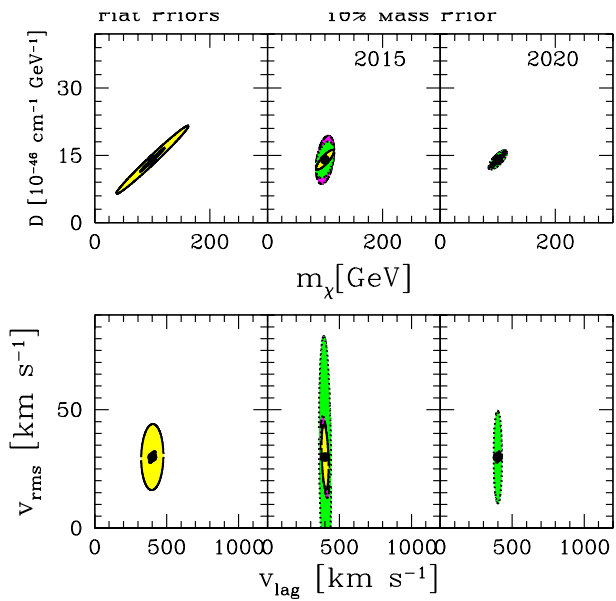}
	\end{center}
	\caption{\label{fig:high}Same particle-physics parameters as Figs. \ref{fig:mA} and \ref{fig:vel} but with $v_{\mathrm{lag}} = 400\hbox{ km s}^{-1}$ and $v_{\mathrm{rms}} = 30\hbox{ km s}^{-1}$.  Ellipse colors have the same meanings as before.}
\end{figure}

In Figs.~\ref{fig:low}~and~\ref{fig:high}, I show 1$\sigma$ uncertainties for different, ``streamlike" velocity parameters but with the same particle-physics parameters as in Figs. \ref{fig:mA} and \ref{fig:vel}.  If $v_{\mathrm{lag}} = v_{\mathrm{rms}} = 30\hbox{ km s}^{-1}$, most nuclear recoils are low in energy, so experiments such as XENON1T or LUX, which have low energy thresholds and heavy target nuclei, are necessary to characterize the velocity and particle-physics parameters, although high-threshold experiments are not constraining.  In Fig.~\ref{fig:low}, the 2015-era experiments do a better job of constraining parameters than the 2020-era experiments because of the lower threshold of XENON1T, although constraints on $D$ are still weak even with the LHC mass prior.  However, the mass prior significantly improves constraints in $v_{\mathrm{rms}}$.

In Fig.~\ref{fig:high}, I show constraints assuming $v_{\mathrm{lag}} = 400\hbox{ km s}^{-1}$ and $v_{\mathrm{rms}} = 30\hbox{ km s}^{-1}$.  In this case, the typical nuclear recoil is large [e.g., Eq. (\ref{eq:vmin})], so experiments with large analysis windows (typical of germanium and argon experiments) are best able to constrain parameters.  Without mass priors, the degeneracy between $m_\chi$ and $D$ is large for 2015-era experiments, but the velocity parameters are better constrained.  For the 2020-era experiments, the uncertainties on the velocity parameters are tiny even without mass priors.  Parameters constraints improve significantly with the LHC mass prior.

There are three takeaway points from this initial study.  (1) It is possible to glean both particle-physics and astrophysics inferences about dark matter by analyzing experiments together if a signal is seen in at least one.  (2) It is only possible to do this by doing a joint analysis of the experiments.  The degeneracies in each experiment are broken by the other experiments.  Thus, having many different dark-matter experiments is a \emph{necessary} condition for being able to extract both particle-physics and astrophysics properties of dark matter from data.  (3) Dark-matter experiments are the \emph{only} probes of the local WIMP DF.  As we come to understand in simulations what drives the evolution of the WIMP DF, we can learn something about the evolutionary history of the Milky Way from the local WIMP DF as unveiled by these experiments.

The joint analysis of dark-matter data sets can and should be used to extract the most information possible about both the particle physics and astrophysics of dark matter from direct-detection experiments and other types of experiments and observations sensitive to dark matter.  This is particularly important because the data sets expected in the next decade will be vast and diverse.  By using robust joint analysis methods, we will be able to explore the complementarity of ongoing and upcoming experimental and observational programs to determining the nature of dark matter.  

\begin{acknowledgments}
I thank Marc Kamionkowski and Sunil Golwala for helpful discussions.  This research was supported by the Gordon and Betty Moore Foundation.  Parts of this work grew out of discussions during a mini study on ``Shedding Light on the Nature of Dark Matter'' at the W. M. Keck Institute for Space Studies.
\end{acknowledgments}


\begin{thebibliography}{24}
\expandafter\ifx\csname natexlab\endcsname\relax\def\natexlab#1{#1}\fi
\expandafter\ifx\csname bibnamefont\endcsname\relax
  \def\bibnamefont#1{#1}\fi
\expandafter\ifx\csname bibfnamefont\endcsname\relax
  \def\bibfnamefont#1{#1}\fi
\expandafter\ifx\csname citenamefont\endcsname\relax
  \def\citenamefont#1{#1}\fi
\expandafter\ifx\csname url\endcsname\relax
  \def\url#1{\texttt{#1}}\fi
\expandafter\ifx\csname urlprefix\endcsname\relax\def\urlprefix{URL }\fi
\providecommand{\bibinfo}[2]{#2}
\providecommand{\eprint}[2][]{\url{#2}}


\bibitem[{\citenamefont{{Komatsu} et~al.}(2010)\citenamefont{{Komatsu},
  {Smith}, {Dunkley}, {Bennett}, {Gold}, {Hinshaw}, {Jarosik}, {Larson},
  {Nolta}, {Page} et~al.}}]{komatsu2010}
\bibinfo{author}{\bibfnamefont{E.}~\bibnamefont{{Komatsu}}}
  \bibnamefont{et~al.}, \bibinfo{journal}{arXiv:1001.4538}.

\bibitem[{\citenamefont{{Alcock} et~al.}(1998)\citenamefont{{Alcock},
  {Allsman}, {Alves}, {Ansari}, {Aubourg}, {Axelrod}, {Bareyre}, {Beaulieu},
  {Becker}, {Bennett} et~al.}}]{alcock1998}
\bibinfo{author}{\bibfnamefont{C.}~\bibnamefont{{Alcock}}}
  \bibnamefont{ et~al.}, \bibinfo{journal}{\apj} \textbf{\bibinfo{volume}{499}},
  \bibinfo{pages}{L9} (\bibinfo{year}{1998}); \textbf{\bibinfo{volume}{542}},
  \bibinfo{pages}{281} (\bibinfo{year}{2000}); \textbf{\bibinfo{volume}{550}},
  \bibinfo{pages}{L169} (\bibinfo{year}{2001}); \bibinfo{author}{\bibfnamefont{P.}~\bibnamefont{{Tisserand}}}\bibnamefont{~et~al.}, \bibinfo{journal}{Astron. Astrophys.}
  \textbf{\bibinfo{volume}{469}}, \bibinfo{pages}{387} (\bibinfo{year}{2007}).

\bibitem[{\citenamefont{{Dodelson} and {Widrow}}(1994)}]{dodelson1994}
\bibinfo{author}{\bibfnamefont{S.}~\bibnamefont{{Dodelson}}} \bibnamefont{and}
  \bibinfo{author}{\bibfnamefont{L.~M.} \bibnamefont{{Widrow}}},
  \bibinfo{journal}{Phys. Rev. Lett.} \textbf{\bibinfo{volume}{72}},
  \bibinfo{pages}{17} (\bibinfo{year}{1994}); \bibinfo{author}{\bibfnamefont{G.}~\bibnamefont{{Jungman}}},
  \bibinfo{author}{\bibfnamefont{M.}~\bibnamefont{{Kamionkowski}}},
  \bibnamefont{and} \bibinfo{author}{\bibfnamefont{K.}~\bibnamefont{{Griest}}},
  \bibinfo{journal}{Phys. Rep.} \textbf{\bibinfo{volume}{267}},
  \bibinfo{pages}{195} (\bibinfo{year}{1996}); \bibinfo{author}{\bibfnamefont{D.~N.} \bibnamefont{{Spergel}}}
  \bibnamefont{and} \bibinfo{author}{\bibfnamefont{P.~J.}
  \bibnamefont{{Steinhardt}}}, \bibinfo{journal}{Phys. Rev. Lett.}
  \textbf{\bibinfo{volume}{84}}, \bibinfo{pages}{3760} (\bibinfo{year}{2000}); \bibinfo{author}{\bibfnamefont{H.-C.} \bibnamefont{{Cheng}}},
  \bibinfo{author}{\bibfnamefont{J.~L.} \bibnamefont{{Feng}}},
  \bibnamefont{and} \bibinfo{author}{\bibfnamefont{K.~T.}
  \bibnamefont{{Matchev}}}, \bibinfo{journal}{Phys. Rev. Lett.}
  \textbf{\bibinfo{volume}{89}}, \bibinfo{pages}{211301}
  (\bibinfo{year}{2002}); \bibinfo{author}{\bibfnamefont{J.~L.} \bibnamefont{{Feng}}},
  \bibinfo{author}{\bibfnamefont{S.}~\bibnamefont{{Su}}}, \bibnamefont{and}
  \bibinfo{author}{\bibfnamefont{F.}~\bibnamefont{{Takayama}}},
  \bibinfo{journal}{\prd} \textbf{\bibinfo{volume}{70}},
  \bibinfo{pages}{063514} (\bibinfo{year}{2004}); \bibinfo{author}{\bibfnamefont{J.~L.} \bibnamefont{{Feng}}} \bibnamefont{and}
  \bibinfo{author}{\bibfnamefont{J.}~\bibnamefont{{Kumar}}},
  \bibinfo{journal}{Phys. Rev. Lett.} \textbf{\bibinfo{volume}{101}},
  \bibinfo{pages}{231301} (\bibinfo{year}{2008}).

\bibitem[{\citenamefont{{Goodman} and {Witten}}(1985)}]{goodman1985}
\bibinfo{author}{\bibfnamefont{M.~W.} \bibnamefont{{Goodman}}}
  \bibnamefont{and} \bibinfo{author}{\bibfnamefont{E.}~\bibnamefont{{Witten}}},
  \bibinfo{journal}{\prd} \textbf{\bibinfo{volume}{31}}, \bibinfo{pages}{3059}
  (\bibinfo{year}{1985}); \bibinfo{author}{\bibfnamefont{I.}~\bibnamefont{{Wasserman}}},
  \bibinfo{journal}{\prd} \textbf{\bibinfo{volume}{33}}, \bibinfo{pages}{2071}
  (\bibinfo{year}{1986}).

\bibitem[{\citenamefont{{Ahmed} et~al.}(2009)}]{cdms2009}
\bibinfo{author}{\bibfnamefont{Z.}~\bibnamefont{{Ahmed}}} \bibnamefont{et~al.},
  \bibinfo{journal}{Phys. Rev. Lett.} \textbf{\bibinfo{volume}{102}},
  \bibinfo{pages}{011301} (\bibinfo{year}{2009}).

\bibitem[{\citenamefont{{Angle} et~al.}(2008{\natexlab{a}})}]{angle2008}
\bibinfo{author}{\bibfnamefont{J.}~\bibnamefont{{Angle}}} \bibnamefont{et~al.},
  \bibinfo{journal}{Phys. Rev. Lett.} \textbf{\bibinfo{volume}{100}},
  \bibinfo{pages}{021303} (\bibinfo{year}{2008}{\natexlab{a}}); 
  \textbf{\bibinfo{volume}{101}},
  \bibinfo{pages}{091301} (\bibinfo{year}{2008}{\natexlab{b}}); \bibinfo{author}{\bibnamefont{{The CDMS Collaboration}}},
  \bibinfo{journal}{arXiv:0912.3592}  (\bibinfo{year}{2009}).

\bibitem[{\citenamefont{{Golwala}}(2010)}]{gaitskell2008}
\bibinfo{author}{\bibfnamefont{S.}~\bibnamefont{{Golwala}}}
  (\bibinfo{year}{2010}),
  \bibinfo{note}{http://www.astro.caltech.edu/\~{}gol- \\wala/talks/20100204GolwalaCaltechWeb.pdf}.

\bibitem[{\citenamefont{{Strigari} and {Trotta}}(2009)}]{strigari2009}
\bibinfo{author}{\bibfnamefont{L.~E.} \bibnamefont{{Strigari}}}
  \bibnamefont{and} \bibinfo{author}{\bibfnamefont{R.}~\bibnamefont{{Trotta}}},
  \bibinfo{journal}{arXiv:0906.5361}.


\bibitem[{\citenamefont{{Binney} and {Tremaine}}(2008)}]{binney2008}
\bibinfo{author}{\bibfnamefont{J.}~\bibnamefont{{Binney}}} \bibnamefont{and}
  \bibinfo{author}{\bibfnamefont{S.}~\bibnamefont{{Tremaine}}},
  \emph{\bibinfo{title}{{Galactic Dynamics}}} (\bibinfo{publisher}{Princeton University Press, Princeton, NJ}, \bibinfo{year}{2008}).

\bibitem[{\citenamefont{{Read} et~al.}(2009)\citenamefont{{Read}, {Mayer},
  {Brooks}, {Governato}, and {Lake}}}]{read2009}
\bibinfo{author}{\bibfnamefont{J.~I.} \bibnamefont{{Read}}}
  \bibnamefont{et~al.},
  \bibinfo{journal}{Mon. Not. R. Astron. Soc.}
  \textbf{\bibinfo{volume}{397}}, \bibinfo{pages}{44} (\bibinfo{year}{2009}); \bibinfo{author}{\bibfnamefont{C.~W.} \bibnamefont{{Purcell}}},
  \bibinfo{author}{\bibfnamefont{J.~S.} \bibnamefont{{Bullock}}},
  \bibnamefont{and}
  \bibinfo{author}{\bibfnamefont{M.}~\bibnamefont{{Kaplinghat}}},
  \bibinfo{journal}{\apj} \textbf{\bibinfo{volume}{703}}, \bibinfo{pages}{2275}
  (\bibinfo{year}{2009}); \bibinfo{author}{\bibfnamefont{T.}~\bibnamefont{{Bruch}}},
  \bibinfo{author}{\bibfnamefont{J.}~\bibnamefont{{Read}}},
  \bibinfo{author}{\bibfnamefont{L.}~\bibnamefont{{Baudis}}}, \bibnamefont{and}
  \bibinfo{author}{\bibfnamefont{G.}~\bibnamefont{{Lake}}},
  \bibinfo{journal}{\apj} \textbf{\bibinfo{volume}{696}}, \bibinfo{pages}{920}
  (\bibinfo{year}{2009}).

\bibitem[{\citenamefont{{Green}}(2003)}]{green2003}
\bibinfo{author}{\bibfnamefont{A.~M.} \bibnamefont{{Green}}},
  \bibinfo{journal}{Phys. Rev. D} \textbf{\bibinfo{volume}{68}},
  \bibinfo{pages}{023004} (\bibinfo{year}{2003}); 
  \bibinfo{journal}{J. Cosmol. Astropart. Phys.}
  \textbf{\bibinfo{volume}{7}}, \bibinfo{pages}{5} (\bibinfo{year}{2008}).


\bibitem[{\citenamefont{{Drees} and {Shan}}(2007)}]{drees2007}
\bibinfo{author}{\bibfnamefont{M.}~\bibnamefont{{Drees}}} \bibnamefont{and}
  \bibinfo{author}{\bibfnamefont{C.-L.} \bibnamefont{{Shan}}},
  \bibinfo{journal}{J. Cosmol. Astropart. Phys.} \textbf{\bibinfo{volume}{6}}, \bibinfo{pages}{11}
  (\bibinfo{year}{2007}).

\bibitem[{\citenamefont{{Sciolla} et~al.}(2008)}]{sciolla2008}
\bibinfo{author}{\bibfnamefont{D.}~\bibnamefont{{Santos}}}
  \bibnamefont{et~al.}, \bibinfo{journal}{J. Phys. Conf. Ser.}
  \textbf{\bibinfo{volume}{65}}, \bibinfo{pages}{012012}
  (\bibinfo{year}{2007}); \bibinfo{author}{\bibfnamefont{H.}~\bibnamefont{{Nishimura}}}
  \bibnamefont{et~al.}, \bibinfo{journal}{J. Phys. Conf. Ser.}
  \textbf{\bibinfo{volume}{120}}, \bibinfo{pages}{042025}
  (\bibinfo{year}{2008}); \bibinfo{author}{\bibfnamefont{G.}~\bibnamefont{{Sciolla}}}
  \bibnamefont{et~al.}, \bibinfo{note}{arXiv:0805.2431}; \bibinfo{author}{\bibfnamefont{M.}~\bibnamefont{{Pipe}}}
  \bibinfo{author}{\bibnamefont{{(Drift Collaboration)}}},
  \bibinfo{journal}{J. Phys. Conf. Ser.} \textbf{\bibinfo{volume}{203}},
  \bibinfo{pages}{012031} (\bibinfo{year}{2010}).

\bibitem[{\citenamefont{{E. Aprile}}(2009)}]{aprile2009b}
\bibinfo{author}{\bibnamefont{{Aprile, E.}}} (\bibinfo{year}{2009}),
  \bibinfo{note}{http://xenon.astro.columbia.edu/\\presentations/Aprile\_SJTU09.%
pdf}.

\bibitem[{\citenamefont{{Akerib} et~al.}(2008)}]{akerib2008}
\bibinfo{author}{\bibfnamefont{D.~S.} \bibnamefont{{Akerib}}}
  \bibnamefont{et~al.}, \bibinfo{journal}{J. Low Temp. Phys.}
  \textbf{\bibinfo{volume}{151}}, \bibinfo{pages}{818} (\bibinfo{year}{2008}).

\bibitem[{\citenamefont{{Szelc}}(2009)}]{szelc2009}
\bibinfo{author}{\bibfnamefont{A.~M.} \bibnamefont{{Szelc}}}, in
  \emph{\bibinfo{booktitle}{4th International Workshop on the Dark Side of the Universe}},
  edited by \bibinfo{editor}{\bibnamefont{{S.~Khalil}}}, \bibinfo{series}{AIP Conf. Proc. No. 1115 (AIP, New York, 2009)}, p. 105.

\bibitem[{\citenamefont{{Baltz} et~al.}(2006)\citenamefont{{Baltz},
  {Battaglia}, {Peskin}, and {Wizansky}}}]{baltz2006}
\bibinfo{author}{\bibfnamefont{E.~A.} \bibnamefont{{Baltz}}},
  \bibinfo{author}{\bibfnamefont{M.}~\bibnamefont{{Battaglia}}},
  \bibinfo{author}{\bibfnamefont{M.~E.} \bibnamefont{{Peskin}}},
  \bibnamefont{and}
  \bibinfo{author}{\bibfnamefont{T.}~\bibnamefont{{Wizansky}}},
  \bibinfo{journal}{\prd} \textbf{\bibinfo{volume}{74}},
  \bibinfo{pages}{103521} (\bibinfo{year}{2006})

\end{thebibliography}

\end{document}